# Structural and evolutionary tunnels of pairwise residue-interaction symmetries connect different structural classes of globular proteins.

**Anirban Banerji**
**Bioinformatics Centre, University of Pune, Pune-411007, Maharashtra, India.**
**Contact Details: anirbanab@gmail.com**

**Abstract**

**Studying all non-redundant proteins in 76 most-commonly found structural domains, we decipher latent patterns that characterize acceptable and unacceptable symmetries in residue-residue interactions in functional proteins. Cutting across the structural classes, a select set of pairwise interactions are found to be universally favored by geometrical and evolutionary constraints; these symmetric interactions are termed 'acceptable' structural and evolutionary tunnels, respectively. An equally small subset of residue-residue interactions, the 'unacceptable' structural and evolutionary tunnels, is found to be universally disliked by structural and evolutionary constraints. Non-trivial overlapping is detected among acceptable structural and evolutionary tunnels, as also among unacceptable structural and evolutionary tunnels. A subset of tunnels is found to have equal relative importance, structurally and evolutionarily, in different structural classes. The MET-MET tunnel is detected to be universally most unacceptable by both structural and evolutionary constraints, whereas the ASP-LEU tunnel was found to be the closest to be called universally most acceptable. Residual populations in structural and evolutionary tunnels are found to be independent of stereochemical properties of individual residues. It is argued with examples that tunnels are emergent features that connect extent of symmetry in residue-residue interactions to the level of quaternary structural organization.**

## 1. Introduction:

Is protein structure organization discreet or continuous? – Many works during the whole of last decade has probed the scope and depth of this question from various perspectives (Shindyalov and Bourne, 2000; Harrison et al., 2002; Hou et al., 2003; Rogen and Fain, 2003; James and Tawfik, 2003; Valas et al., 2009). The present work also attempts to approach this question, but from a different starting point to that of the previous works.

We start with no assumption but with only the basic piece of knowledge that protein native structures embody a delicate balance between local and non-local interactions; where the former, rather independently of the global context of structural organization, ensures the formation of stable secondary structures and the later ensures formation of stable tertiary structure (Freund et al., 1996; Han and Baker, 1996; Dill, 1999). The fact that some small stretches of primary structure may adopt multiple secondary structure conformations in different proteins (viz., the chameleon sequences (Minor and Kim, 1996; Mezei, 1998)), exemplifies the connection between (aforementioned) local interactions and the global framework. But examples are not rules. In this work we attempt to decipher the rules that govern the scheme of connection between patterns in local interactions and principles governing global structural organization. We argue further that by investigating the nature of these rules, one may infer the extent to which protein structural organization is discreet or continuous.

While primary structure of a protein encodes the folding code in some definite but not-yet fully-deciphered manner, it is at the level of secondary structures that we first encounter the hydrogen bonds and various other chemical interactions, which in multifarious forms add on to themselves till the biologically functional quaternary structure is obtained. Protein organization is marked by a spectrum of short and long range interactions (Munoz et al., 1996; Gromiha and Selvaraj, 2001; Kumar and Nussinov, 2002; Di Paola et al., 2013). The root of (possible) continuous nature of protein structural organization, therefore, lies neither in the individual residues themselves, nor in the primary structures, but in the residue-residue interactions, where the residues may or may not be part of the same secondary structure. – This raises more questions than answers; viz.,

**A:** Do there exist deep patterns that underlie the traditional schemes (viz., Karplus and Shakhnovich, 1992; Fiebig and Dill, 1993; Plaxco et al., 1998; Miyazawa and Jernigan, 1999; Mirny and Shakhnovich, 2001; Lu and Skolnick, 2001; Seno et al., 2008; among many others) to describe residue-residue interactions?

**B:** If indeed some general and latent patterns are found beneath contact-order and contact-matrix information, how can these residual interaction patterns be connected to the organizational schemes of biologically functional states of these proteins?

**C:** Do these deep patterns in pairwise interactions truly constitute continuous threads that connect residue-residue interactions to quaternary structures?

**D:** How will the residue-residue interaction data be analyzed from the perspective of protein evolution?

**E:** Can one analyze the commonality and differences among these latent patterns that characterize residue-residue interaction profile in different structural classes?

– Considering all the available non-redundant proteins belonging to 76 commonly found structural domains across four major structural classes, the present work attempts to answer each of the aforementioned questions. Furthermore, it explores the scope and depth of the simple question: though all types of residue-residue interactions are important, looking through the prism of structure evolution, which pairwise interactions are more important than the others within a particular SCOP (Murzin et al., 1995) class and across all the SCOP classes?

To achieve certain globally defined goal at the biochemical pathway level under a specific context, a protein is made to function. The evolutionary-fitness of every aspect of structural organization of any protein can therefore be measured by the accuracy and efficiency with which this structure can function. While residue-residue interactions primarily characterize local (pairwise) interactions in a protein, the evolutionary-fitness of each of these interactions can only be measured by some suitable quantity that connects pairwise interactions to the global scheme of structural organization of a functioning protein. We propose that, symmetry in residue-residue interactions, as measured through a non-linear dynamics based construct, the 'correlation dimension'(CD) (Grassberger and Procaccia, 1983; Takens, 1985), can connect the level of local proximities and preferred nature of local pairwise interactions to the global level of evolutionarily preferred scaffold of interaction sets. Though attempts to study the extent of continuity in protein structural organization isn't new (Presnell and Cohen, 1989; Baldwin and Rose, 1999; Salem et al, 1999; Petrey and Honig, 2000; Szustakowski et al., 2005; Petrey et al., 2009), works on the interplay and trade-offs between local stereochemical and global evolutionary preferences (or constraints), connecting the patterns at the level of residue-residue interactions to the organizational principles at the level of quaternary structures – are not easy to find.

To add another level of complexity to the problem we find that proteins are fractal objects and are not classical (viz., Debye) solids (Banerji and Ghosh, 2009; de Leeuw et al., 2009). This is relevant, because CD is special type of fractal dimension, particularly useful to obtain an objective idea about the symmetry in the distribution of the pairwise-interactions. Whereby, starting merely with a qualitative idea about the distribution of residue-residue interactions in protein space, CD quantifies the extent of underlying symmetry in each of these interactions by neither imposing Euclidean constructs on proteins nor by studying protein properties as deviations from some (illusory)ideality, but simply by treating proteins as complex systems. Though CD is routinely employed in studies of chaotic dynamics (to determine dimensions of (strange) attractors), its use in protein studies (and biological macromolecules in general) is rare. While it was used to investigate morphology of protein exterior (Choi and Lee, 2000), backbone properties (Lee, 2008) and distribution of interior biophysical properties (Tejera et al., 2009; Banerji and Ghosh, 2011) – finding other works that attempted to investigate proteins with CD, proved difficult.

A possible starting point to investigate the common evolutionary platform that allows structural equivalence among continuous (and not discreet) building blocks of folded polypeptides across proteins belonging to different structural classes could have been expected to be based on the energetic considerations (Finkelstein and Ptitsyn, 1987; Chothia and Finkelstein, 1990) too. However, number of low energy structures in a protein is enormous and many locally-global minima in energy landscape are known to compete with the ground state of a functional protein. Moreover, the thermodynamically-stable global minima of a protein may not necessarily correspond to its (evolutionally-approved) functional state. Also, since only a selected few of the low energy structures turn out to be highly designable ones (Li et al., 1998) and since only the highly designable structures are found to possess ("protein-like") geometrical regularities (Li et al., 1998) – an investigation through purely energetic considerations may fail to probe the present set of questions in hand. Finally, since it is difficult to construct a general way to connect the energetic state of a stretch of polypeptide to the symmetry of residue-residue interactions in it, the energy-based approach was not adopted.

## 2. Materials and Methods:

*2.1.Materials:* It is known that the total number of distinct folding patterns isn't infinite but limited, to the extent that only a few dozen folding patterns is found to account for half of all the known protein structures (Voet and Voet, 2010). This finding demonstrates the presence of common templates of protein folding patterns, which act as structural and evolutionary "sinks" (Nelson and Cox, 2008). Thus, in order to investigate the nature of structural and evolutionary sinks in folding patterns, we did not require studying the entire universe of known protein structures; instead, we studied all the non-redundant proteins belonging to 76 commonly found SCOP folds, spanning across 4 major structural classes of globular proteins, all-$\alpha$, all-$\beta$, $\alpha/\beta$ and $\alpha + \beta$. To capture the subtlety of residue-residue interactions in their finer details, only structures with high resolution ($\leq 2.5$ Å) were considered for this study. Additionally, it was made sure that none of the proteins contained any "disordered" region (as defined in (Dunker et al., 2001)) in them.

Principal purpose of the study was to detect and probe the underlying patterns that not only act as structural templates for stable folding but also ensure support to biological functions to survive evolutionary constraints. Hence, functional forms of proteins, viz. proteins in their quaternary structures were considered. The coordinate file for asymmetric unit of a protein merely informs about the fraction of the crystallographic unit cell that has no crystallographic symmetry. These may contain artifacts of crystallization (Valdar and Thornton, 2001) and in general, may represent the biologically functional state of the protein rather infrequently. Biological unit of a protein, as provided by the PDB (Berman et al., 2003), presents the entire information about biologically relevant multimeric state of the functional protein. Thus, to suit the requirement of the study, biological units of all the proteins (instead of their asymmetric units) were considered.

Upon downloading the biological unit of a protein, the SCOP domain characterization of each chain (at times segment of a chain) of the protein was obtained. Residue-residue interactions were studied upon these segregated SCOP-domains.

*2.2. Methodology:* Three dimensional structure of a protein originates from its primary structure. For every protein, synthesis of each element (viz. residue) of its primary structure takes place in a definite sequential order from its N-terminal to C-terminal. Considering the primary structure as a special kind of mathematical series, one may choose to view the position of each residue in the folded protein as an embedding $\{r\}_{i=1}^{N}$ of the primary structure in the space of folded protein. In this description of residual position, the 'correlation function' $C_N(\varepsilon)$ is defined as:

$$C_N(\varepsilon) = \binom{N}{2}^{-1} \sum_{0 \leq i < j \leq N} H\left( \left\| \vec{r_i} - \vec{r_j} \right\| < \varepsilon \right) \qquad \text{Eq}^n\text{ - 1}$$

where H(X) is the Heaviside function whose value is 1 if the condition X is satisfied, 0 otherwise; and $\|.\|$ denotes the distance induced by the selected norm within the structural space of folded protein. The sum $\sum_i H(\|\vec{r_i} - \vec{r_j}\| < \varepsilon)$ is the number of residues (represented by the coordinates of their respective $C^\beta$ atoms ($C^\alpha$ for GLY)) within a distance $\varepsilon$ of each other (numerically, $C^\beta - C^\beta \leq 8$Å). The correlation dimension (CD) can then be defined as:

$$CD = \lim_{\varepsilon \to 0} \lim_{N \to \infty} \frac{\log C_N(\varepsilon)}{\log(\varepsilon)} \qquad \text{Eq}^n\text{ - 2}$$

Amongst 20 residues of a protein, there can be 210 [=(190(= $^{20}_2C$)+20)] types of possible residue-residue interactions. Thus 210 CD magnitudes were calculated for each protein considered. These magnitudes were averaged for SCOP folds, which in turn, were averaged to the level of SCOP classes.

The reason for having the (somewhat uncommon) normalization factor in correlation function $C_N(\varepsilon)$ is that, with this normalization we don't define $C_N(\varepsilon)$ as an estimate of the expected number of residues within a radius $\varepsilon$ of any given residue; instead, $C_N(\varepsilon)$ becomes an estimate of the probability that two residues chosen at random within a protein reside within a distance threshold $\varepsilon$. Difference between expectation and probability is only a constant of proportionality, if, the residues are distributed uniformly within a protein. Of course, in such a case, the constant of proportionality vanishes in the limit of eq$^n$-2. The reason for choosing the probability rather than the expectation is that, with such a definition, the concept of dimension still makes sense and in fact it generalizes to cases where the sample spatial distribution of residues are not distributed uniformly (Judd, 1992), which is exactly what the general case of residue distribution within proteins is.

### 3. Results:

Individual symmetries of each 210 residue-residue interactions, was calculated for all non-redundant proteins in 76 most-commonly found SCOP folds. Magnitude of each of these pairwise-interactions symmetries, averaged at the level of each of the SCOP folds, is provided in **Suppl. Mat.-1**. Intuitively, one may expect that all these 210 pairwise interactions are equally symmetric, all are equally favored by proteins belonging to different structural classes and the structural classes themselves do not favor or disfavor the extent of symmetry possessed by any of these 210 interactions. However, we found that pairwise interactions are accommodated within structural classes in a highly biased manner. The extent of symmetry of pairwise interactions in any structural class is found to demonstrate highly skewed distribution and only a tiny fraction of the 210 pairwise interactions are found to possess high-degree of symmetry. More importantly, the interactions that are detected to be highly symmetric (and favored, structurally), are found to be highly symmetric (and favored, evolutionarily) across all the structural classes! On the other hand, nature is found to be even more decisive in disliking certain pairwise interactions. Whereby, pairwise interactions that recorded low extent of symmetry are found to be more in number than pairwise interactions with prominent symmetries. Moreover, interactions that recorded low extent of symmetry in any of the four structural classes are found to record low extent of symmetry in all the structural classes; in fact they are found to be disliked both structurally and evolutionarily, universally. Among all the 210 pairwise interactions in all four structural classes, GLU-ASP interaction in $\alpha/\beta$ proteins is found to be the most-symmetric residue-residue interaction recording $CD^{\alpha/\beta}_{GLU-ASP} = 2.555$ . This implies that effect of symmetry of self-similarity due to GLU-ASP interactions in $\alpha/\beta$ proteins fills protein space to more than half of its radial extent; had it filled the 3-D space of protein completely, CD magnitude of it would have been 3.00. On the other extreme, the extent of symmetry of TRP-TRP interactions in all-$\alpha$ proteins is found to be the minimum, $CD^{all-\alpha}_{TRP-TRP} = 0.764$, implying that the effect of symmetry of self-similarity due to TRP-TRP interactions in $(all-\alpha)$ class of proteins is minimally space-filling (CD=0.0 being the biophysical impossibility of any type of pairwise interaction with no space-filling effect due to its symmetry at all). These results are presented in details in **Section-3.1**.

Evolutionary constraints have innately context-dependent requirements in them. Remarkably, it was found that not only the geometric constraints on the organization of structural classes, but the context-specific evolutionary constraints on the functional state of proteins – also prefer only a selected subset of pairwise interaction symmetries and do not like most of the 210 possibilities. Thus, not only the structural/geometrical constraints but also the evolutionary constraints on protein structures are found to demonstrate a highly biased scheme of mapping the extent of symmetry in pairwise residual interactions to structural classes. These results are presented in **Section-3.2**.

Equivalence of relative importance of any interaction symmetry in two, three or all four structural classes is an extremely improbable event to conceive. Nevertheless, many residue-residue interaction symmetries were detected to have equal relative importance, both structurally and evolutionarily, across two or more structural classes! These results are presented in **Section-3.3**. Commonalities and contrasting features among structurally-preserved and evolutionarily-preserved pairwise interactions are dissected and discussed in **Section-3.4**.

Notably, it was found that pairwise residual interactions that have recorded maximum symmetry in all-$\alpha$ or all-$\beta$ structural classes are not at all dependent upon Chou-Fasman propensity of individual amino acids [**CF REF**] in $\alpha$-helix or $\beta$-sheet respectively! These results are presented in **Section-3.5**. The failure to explain structurally and evolutionarily favored interaction-symmetries with stereochemical properties of single residues, hinted at the emergent nature of these interaction-symmetries, which could only be detected at the level of quaternary structural organization. Referring to concrete examples of residual properties, we argued (in **Section-3.6**), why, symmetry in pairwise interactions should be considered as examples of 'strong-emergence'(Yaneer Bar-Yam, 2004) in the paradigm of protein structure studies.

*3.1. The Structural Tunnels:* Any structural class is characterized by its unique set of stereochemical and stability constraints that distinctly distinguish the geometry-dependent mechanical load distribution scheme in its scaffold. One may intuitively expect that differences in structural/stereochemical/geometrical constraints across the structural domains, in some way or the other, will be reflected in differences in the basis of all organizational platforms, viz. in the residue-residue interaction profiles. Whereby, *one may expect that each structural class will be characterized by a very distinct set of pairwise interactions of its own, which are either categorically favored or categorically disfavored*. Such trademark pairwise interactions for a structural class are expected to be characterized by archetypal symmetry profiles that are unique to the structural class the interactions belong to. Whereby, out of 210 possibilities, the most-symmetric 10 pairwise interactions in any structural class are expected to represent the distinctive set of symmetries that structural/stereochemical/geometrical constraints in that structural class can permit and conserve. An occurrence of repetition of any of the 10 most-symmetric pairwise interactions recorded for one structural class, in another, may therefore be considered (structurally) improbable.

In reality however, we found that, first, no such trademark-set of symmetric pairwise interactions characterizes any of the structural classes. It is found that high degree of symmetry in pairwise interactions was observed only for a small-subset of 210 possibilities and elements of this small-subset are shared by all the structural classes; whereby no archetypal residual interaction symmetry could be found for any of the structural classes. Second, (elaborating on first), irrespective of the differences in the multi-parametric structural constraints of the structural classes, a total of only 5 pairwise interactions are found to feature consistently as the 3 most-symmetric pairwise interactions across all four structural classes. If nature was unbiased, the total number of 3-most-symmetric pairwise interactions across four structural classes would have been 535933200168609058816000O($=(^{210}_{3}C)^4$); instead, only 5 (ARG-LYS, GLU-ASP, ARG-ARG, PHE-GLU, ASP-LEU) were found. Among these, the ARG-LYS interaction was found to be present consistently as the most symmetric pairwise interactions in three (all-$\alpha$, all-$\beta$, $\alpha+\beta$) out of four structural classes, which are marked with significantly different set of stereochemical, geometric and stability constraints.

Trends observed for 5-most-symmetric pairwise interactions extend for the 10-most-symmetric pairwise interactions across four structural classes too. If nature was unbiased, the total number of 10-most-symmetric pairwise interactions across four structural classes would have been astronomically high : $(^{210}_{10}C)^4$ ; instead, merely 22 were detected. Out of this select set of 22, a highly exclusive subset of 9 interaction symmetries (<

5% of 210 possibilities) are found to populate the ranks of 10-most-symmetric pairwise interactions in two or more structural classes. More importantly, suggesting bias-within-bias, 4 out of the 9 aforementioned symmetric interactions are found to be consistently present amongst 10-most-favored symmetric interactions across all structural classes. The skewed distribution of these 22 interactions is represented by:

a. 4 'favored' pairwise interaction symmetries across all 4 structural classes (ARG-LYS, GLU-ASP, PHE-GLU, ASP-LEU)
b. 1 'favored' pairwise interaction symmetry across 3 structural classes (LYS-PHE, in all-$\alpha$, all-$\beta$, $\alpha/\beta$)
c. 4 'favored' pairwise interaction symmetries across 2 structural classes (PHE-ASP(all-$\beta$, $\alpha + \beta$), ASP-VAL(all-$\beta$, $\alpha + \beta$), ASP-ILE(all-$\beta$, $\alpha + \beta$), ARG-PHE(all-$\beta$, $\alpha/\beta$)
and
d. 13 'favored' pairwise interaction symmetries; those are specific to any one of the structural classes.

In order to not miss out any 'favored' residual interaction symmetries, we chose a more inclusive criterion; any pairwise interaction that occurs within the 25 most-symmetric pairwise interactions for more than one structural class, is defined as acceptable 'Structural Tunnel'(ST). Acceptable-STs demonstrate nature's preference for a limited set of pairwise interaction symmetries that are consistently conserved across different milieu of structural-and-geometric constraints. Acceptable-STs comprise of <5% of 210 pairwise interaction possibilities (**Table-1**). On the other end of the spectrum, we detected existence of STs that are universally disliked across structural classes, registering low extent of interactions symmetries; these are named the 'non-acceptable-STs' (**Table-2**). The non-acceptable-STs are comparable in their number to the acceptable-STs; where the later is comprised of 9 interaction symmetries, the former is comprised of 12 interactions.

**Table-1: Acceptable Structural Tunnels**

| ALL-ALPHA | | ALL-BETA | | ALPHA/BETA | | ALPHA+BETA | |
|---|---|---|---|---|---|---|---|
| RES-RES | CD | RES-RES | CD | RES-RES | CD | RES-RES | CD |
| arg-lys | 2.340389 | arg-lys | 2.421 | glu-asp | 2.555048 | arg-lys | 2.435679 |
| glu-asp | 2.261833 | phe-glu | 2.417963 | arg-lys | 2.529714 | glu-asp | 2.42425 |
| arg-arg | 2.251167 | asp-leu | 2.416741 | asp-leu | 2.510571 | asp-leu | 2.406464 |
| phe-glu | 2.242 | phe-asp | 2.399296 | phe-glu | 2.499524 | asp-ile | 2.378893 |
| lys-lys | 2.226111 | glu-asp | 2.398556 | lys-phe | 2.495286 | phe-asp | 2.377393 |
| his-lys | 2.220333 | glu-leu | 2.385815 | arg-ile | 2.494095 | asp-val | 2.374893 |
| lys-phe | 2.2075 | asp-val | 2.383074 | asn-val | 2.486048 | glu-ile | 2.374071 |
| ser-glu | 2.202222 | lys-phe | 2.383 | lys-ile | 2.481143 | phe-glu | 2.373 |
| asp-leu | 2.197611 | asp-ile | 2.366778 | ala-asp | 2.473143 | glu-glu | 2.364036 |
| glu-gly | 2.196944 | arg-phe | 2.361852 | arg-phe | 2.469476 | ser-lys | 2.350321 |
| ala-pro | 2.195 | asn-leu | 2.356259 | arg-val | 2.46581 | arg-phe | 2.346286 |
| ala-his | 2.186222 | glu-glu | 2.353259 | phe-asp | 2.464571 | his-lys | 2.343429 |
| asn-glu | 2.178389 | lys-lys | 2.346444 | gln-val | 2.464524 | asn-val | 2.341571 |
| lys-tyr | 2.177111 | asn-ile | 2.345444 | glu-ile | 2.46419 | ala-asp | 2.341071 |
| thr-arg | 2.1705 | glu-ile | 2.342037 | glu-gly | 2.464048 | arg-leu | 2.337036 |
| ser-lys | 2.168389 | lys-leu | 2.340333 | glu-val | 2.461905 | tyr-glu | 2.334286 |
| pro-glu | 2.164611 | asn-val | 2.336111 | pro-lys | 2.458905 | ser-glu | 2.334214 |
| tyr-glu | 2.164333 | asn-phe | 2.335 | asp-ile | 2.457762 | asn-leu | 2.333821 |
| gln-arg | 2.162722 | arg-ile | 2.326889 | asn-leu | 2.456143 | pro-lys | 2.332464 |

| | | | | | | |
|---|---|---|---|---|---|---|
| arg-phe | 2.161444 | pro-ile | 2.324926 | lys-val | 2.452905 | lys-phe | 2.33075 |
| pro-leu | 2.158333 | ser-glu | 2.320889 | pro-glu | 2.452429 | glu-gly | 2.329821 |
| glu-val | 2.157222 | pro-lys | 2.320741 | asp-val | 2.450571 | pro-glu | 2.329464 |
| lys-val | 2.152389 | glu-val | 2.316667 | arg-leu | 2.446619 | arg-his | 2.328964 |
| pro-lys | 2.150778 | lys-ile | 2.316333 | pro-arg | 2.445 | glu-leu | 2.32525 |
| pro-ile | 2.1495 | arg-arg | 2.314407 | asn-ile | 2.443619 | glu-val | 2.319679 |

**Table-1 Legend:** 25 most-symmetric residue-residue interactions are enlisted above. An interaction that registers itself among 25 most-symmetric (and therefore, most-favored) interactions in more than one structural class, is termed as an 'acceptable structural tunnel'. Acceptable structural tunnels those are favored in all four structural classes are: ARG-LYS, GLU-ASP, PHE-GLU, ASP-LEU, LYS-PHE, ARG-PHE, GLU-VAL, PRO-LYS. Acceptable structural tunnels favored in three structural classes are: PHE-ASP, ASP-ILE, ASP-VAL, ASN-VAL, GLU-ILE, SER-GLU, GLU-GLY, ASN-LEU, PRO-GLU. Acceptable structural tunnels favored in two structural classes are: ARG-ARG, LYS-LYS, HIS-LYS, GLU-LEU, ARG-ILE, LYS-ILE, ALA-ASP, GLU-GLU, SER-LYS, ASN-ILE, TYR-GLU, PRO-ILE, LYS-VAL.

## Table-2: Non-Acceptable Structural Tunnels

| ALL-ALPHA | | ALL-BETA | | ALPHA/BETA | | ALPHA+BETA | |
|---|---|---|---|---|---|---|---|
| **RES-RES** | **CD** | **RES-RES** | **CD** | **RES-RES** | **CD** | **RES-RES** | **CD** |
| tyr-trp | 1.7935 | met-ile | 2.017111 | met-his | 2.110762 | phe-ile | 2.000679 |
| phe-trp | 1.787222 | met-phe | 2.013222 | gln-gln | 2.110762 | cys-phe | 2.000464 |
| ala-cys | 1.783167 | cys-pro | 2.01 | lys-glu | 2.108667 | tyr-tyr | 1.999321 |
| phe-leu | 1.783167 | phe-trp | 2.005 | arg-glu | 2.104048 | cys-val | 1.995393 |
| phe-phe | 1.779722 | gln-trp | 1.999815 | cys-ile | 2.100619 | phe-leu | 1.993821 |
| trp-ile | 1.779667 | phe-leu | 1.976519 | cys-phe | 2.095524 | tyr-trp | 1.97275 |
| ile-ile | 1.776667 | phe-ile | 1.972519 | cys-pro | 2.07019 | val-ile | 1.962786 |
| leu-ile | 1.764167 | cys-tyr | 1.971815 | phe-phe | 2.061571 | pro-trp | 1.957179 |
| met-phe | 1.762 | his-his | 1.965 | cys-val | 2.06081 | trp-ile | 1.952714 |
| met-tyr | 1.741056 | cys-phe | 1.950519 | val-leu | 2.049476 | leu-ile | 1.952179 |
| leu-leu | 1.736056 | leu-ile | 1.94037 | leu-ile | 2.048714 | val-leu | 1.949929 |
| cys-ile | 1.711 | leu-leu | 1.938926 | leu-leu | 2.024762 | val-val | 1.949286 |
| cys-val | 1.683222 | val-leu | 1.922444 | his-trp | 2.023095 | leu-leu | 1.937643 |
| cys-phe | 1.679389 | ile-ile | 1.906444 | his-his | 2.015 | phe-phe | 1.934071 |
| met-met | 1.656889 | cys-his | 1.899852 | cys-met | 2.011571 | his-trp | 1.926143 |
| cys-tyr | 1.641056 | phe-phe | 1.894519 | cys-tyr | 2.010905 | phe-trp | 1.917214 |
| cys-thr | 1.632 | val-ile | 1.892704 | ile-ile | 1.988571 | ile-ile | 1.912107 |
| cys-pro | 1.565611 | val-val | 1.891333 | cys-his | 1.984667 | met-met | 1.869107 |
| his-trp | 1.562333 | his-trp | 1.888556 | val-val | 1.970381 | cys-his | 1.820571 |
| met-trp | 1.530667 | met-trp | 1.807 | val-ile | 1.968143 | cys-met | 1.803464 |
| cys-his | 1.525889 | met-met | 1.801593 | met-met | 1.929048 | his-his | 1.766107 |
| cys-met | 1.496222 | cys-trp | 1.79737 | met-trp | 1.772095 | met-trp | 1.664179 |
| cys-trp | 1.481833 | cys-met | 1.679111 | trp-trp | 1.74181 | cys-trp | 1.61075 |
| cys-cys | 1.1355 | trp-trp | 1.630481 | cys-trp | 1.716571 | cys-cys | 1.348643 |
| trp-trp | 0.763667 | cys-cys | 1.162037 | cys-cys | 1.624333 | trp-trp | 1.269071 |

**Table-2 Legend:** 25 least-symmetric residue-residue interactions are enlisted above. An interaction that registers itself among 25 least-symmetric (and therefore, least-favored) interactions in more than one structural class, is termed as a 'non-acceptable structural tunnel'. Non-acceptable structural tunnels those are least-favored in all four structural classes are: CYS-PHE, PHE-PHE, ILE-ILE, LEU-ILE, CYS-TRP, LEU-LEU, HIS-TRP, MET-MET, CYS-HIS, CYS-MET, MET-TRP, TRP-TRP, CYS-CYS. Non-acceptable structural tunnels disfavored in three structural classes are: PHE-TRP, CYS-PRO, PHE-LEU, CYS-VAL, VAL-ILE, CYS-TYR, HIS-HIS, VAL-LEU, VAL-VAL. Non-acceptable structural tunnels disfavored in two structural classes are: TYR-TRP, PHE-ILE, MET-PHE, CYS-ILE, TRP-ILE.

To help the representation from this point onward, the number of occurrences of a tunnel in structural classes will be referred to as 'tunnel strength'(#).

To characterize the degree of universal disliking, before considering all 25 least symmetric residual interactions (viz., all the non-acceptable-STs), we concentrate on 3 least-symmetric and 10 least-symmetric interactions. The scale of the disliking-bias can be easily estimated from the fact that if nature was unbiased, the total number of 3 least-symmetric pairwise interactions across four structural classes would have been (as demonstrated before), a staggering $(^{210}_{3}C)^4$. Instead, only 4 pairwise interactions were found to occupy all the 12 slots, viz. slots for 3 least-symmetric pairwise interactions in 4 structural classes. Interactions among CYS and TRP residues were noted to be particularly unacceptable; not only because TRP-TRP and CYS-CYS interaction symmetries were found to be populating consistently across 3-least-symmetric interactions in all four structural classes, but also because CYS-TRP, TRP-TRP and CYS-CYS interactions were recorded as 3 least-ranked interaction symmetries in three (all-$\alpha$, $\alpha/\beta$, $\alpha+\beta$) out of four structural classes.

Interestingly, the distribution of the unacceptable interaction-symmetries across structural classes was found to be significantly less-skewed than that of the distribution of favored-interaction-symmetries. As reported beforehand, within topmost-10 slots of symmetric interactions across all four structural classes, 13 favorable interaction-symmetries were accommodated within just one of the four structural classes, whereas within the least-symmetric 10 interactions, only a trifle 5 unfavorable interaction-symmetries were found to be disliked by a single structural class. This suggests that while nature has identified few interaction-symmetries as favorable for structural/geometric constraints of anyone of a single structural class, it has identified a more comprehensive set of interaction-symmetries that are unfavorable for structural/geometric constraints of more than one structural class. The decisiveness in nature's disliking has resulted in less-skewed distribution of 10 least-symmetric residual interactions to structural classes; it is given by:

a. 4 'disliked' pairwise interaction symmetries across all 4 structural classes (MET-TRP, CYS-TRP, TRP-TRP, CYS-CYS)
b. 3 'disliked' pairwise interaction symmetries across 3 structural classes (CYS-HIS (all-$\alpha$, $\alpha/\beta$, $\alpha+\beta$), MET-MET (all-$\beta$, $\alpha/\beta$, $\alpha+\beta$), CYS-MET (all-$\alpha$, all-$\beta$, $\alpha+\beta$))
c. 5 'disliked' pairwise interaction symmetries across 2 structural classes (CYS-TYR (all-$\alpha$, $\alpha/\beta$), VAL-ILE (all-$\beta$, $\alpha/\beta$), ILE-ILE($\alpha/\beta$, $\alpha+\beta$), VAL-VAL (all-$\beta$, $\alpha/\beta$), HIS-TRP (all-$\alpha$, all-$\beta$))
and
d. 5 'disliked' pairwise interaction symmetries; those are specific to any one of the structural classes.

*3.2. The Evolutionary Tunnels:* The evolutionary tunnel study attempted to answer two simple questions, namely, how many and which residue-residue interactions are most-favored evolutionarily across structural classes? Equally, how many and which residue-residue interactions are least-favored evolutionarily across structural classes?

The innately hierarchical nature of SCOP ensures that proteins within the same evolutionary superfamily are assigned to the same fold group (Harrison et al., 2002). A previous work has assigned evolutionary scale index magnitude to SCOP domains, by performing phylogenetic analyses. The evolutionary scale index magnitude provides a way to measure the age of a fold by numerically describing the time-scale at which this fold appeared, whereby structural domains that appeared early in the course of protein evolution could be distinguished clearly from the structural domains that appeared recently. Evolution judges the usefulness of any structure by observing the efficiency with which it performs the function assigned to it. Elements of structural organization that fail to register as functionally competent are discarded by evolutionary mechanisms. Assuming that evolutionary constraints dictate all the levels of protein structural organization (from the level of the functional form of quaternary structure to every underlying level), if the extent of symmetry of anyone of 210 pairwise-interactions is found to increase during the process of evolution, this pairwise-interaction can be considered to be one favored by the evolutionary constraints. On the other hand, if the extent of symmetry in a pairwise-interaction is found to decrease during evolution, the pairwise-interaction can be considered as an evolutionarily non-favored interaction. For each of the 210 pairwise interactions, a careful analysis of correlation coefficient between the extent of symmetry of this interaction in proteins belonging to any fold and the age of the fold, over the entire range of evolutionary age and entire range of structural domains – appears to be a reasonable strategy to identify the pairwise residual interactions favored by evolution from the ones that fell out of evolutionary preference. But then, one cannot assume the evolutionary constraints to be uniform across the spectrum of structural classes. Hence, the aforementioned correlation coefficients were calculated in a structural class specific manner, over the entire span of evolutionary age.

It was found that, notwithstanding the differences in structural class-specific evolutionary constraints, for certain interactions, the extent of embodied-symmetry are increasing *across* the structural classes; while for certain other interactions, their embodied-symmetry is decreasing, *cutting across* the structural classes. In other words, during the process of evolution, a small subset of 210 pairwise residual interactions are favored to the extent that, notwithstanding the different constraints on the structural class-specific evolutionary fitness landscape, the symmetry profiles of these pairwise interactions were approved to increase; while regardless the type of structural class, extent of symmetry of certain other pairwise interactions were made to decrease. In analogy to the STs, we call the pairwise residual interactions which records increasing/decreasing correlation coefficients with evolutionary scale in more than one structural class to be acceptable/non-acceptable evolutionary tunnels (ETs).

Someone remarked "anything that is produced by evolution is bound to be a bit of mess" (- Sydney Brenner, as quoted in (Walsh, 1985)). Since it is difficult to identify all the factors that influence the structural class-specific evolution of protein structures, to take into account our lack of knowledge and other parameters that may have contributed to the "messy" character of evolution, instead of 25 most-symmetric pairwise residual interactions (as considered to detect the acceptable STs), a doubly-inclusive, 50 pairwise residual interactions with highest correlation coefficients were identified as the acceptable-ETs. Similarly, 50 pairwise residual interactions with lowest correlation coefficients were identified as the non-acceptable-ETs. Extent of symmetry in THR-ARG interactions in all-$\alpha$ proteins and that of ALA-ILE interactions in all-$\beta$ proteins could be detected to grow most notably during evolution, registering the correlation coefficients 0.536 and 0.532 respectively, between their CD magnitudes and evolutionary indices of the folds that their host proteins belonged to. In the opposite extreme, the ALA-ALA interaction symmetry followed by ARG-VAL, PRO-PRO, ALA-PHE and ALA-TRP symmetries – in that order, all in $\alpha/\beta$ proteins – showed definite decrements in respective symmetries during the course of evolution. Details of acceptable and non-acceptable ETs are enlisted in **Table-3** and **Table-4**, respectively.

## Table-3: Acceptable Evolutionary Tunnels

| ALL-ALPHA | | ALL-BETA | | ALPHA/BETA | | ALPHA+BETA | |
|---|---|---|---|---|---|---|---|
| **Res-Res** | **CC** | **Res-Res** | **CC** | **Res-Res** | **CC** | **Res-Res** | **CC** |
| thr-arg | 0.536 | ala-ile | 0.532 | thr-phe | 0.44083 | gly-gly | 0.232 |
| gln-arg | 0.501 | gln-leu | 0.482 | lys-asp | 0.34152 | ala-val | 0.197 |
| asn-ile | 0.499 | pro-phe | 0.469 | ala-gln | 0.34056 | thr-val | 0.183 |
| thr-tyr | 0.49 | ser-val | 0.466 | asp-leu | 0.33129 | cys-pro | 0.179 |
| arg-trp | 0.473 | gln-tyr | 0.446 | his-phe | 0.29949 | his-his | 0.172 |
| met-pro | 0.466 | ala-val | 0.444 | pro-asn | 0.29542 | gln-arg | 0.164 |
| pro-gly | 0.459 | phe-gly | 0.437 | lys-val | 0.29162 | leu-gly | 0.16 |
| gln-phe | 0.454 | ser-trp | 0.425 | gln-leu | 0.27054 | asp-leu | 0.16 |
| ala-met | 0.444 | val-gly | 0.424 | pro-tyr | 0.2436 | arg-asp | 0.154 |
| met-thr | 0.441 | thr-leu | 0.42 | lys-leu | 0.23205 | cys-asn | 0.15 |
| ala-his | 0.441 | ala-leu | 0.417 | ala-glu | 0.21146 | ser-trp | 0.146 |
| met-lys | 0.437 | thr-gly | 0.417 | lys-ile | 0.21017 | asn-ser | 0.141 |
| pro-tyr | 0.431 | phe-asp | 0.416 | asn-trp | 0.20121 | cys-ser | 0.13 |
| thr-val | 0.414 | thr-trp | 0.412 | cys-cys | 0.13998 | met-leu | 0.126 |
| asn-val | 0.406 | arg-glu | 0.41 | ala-lys | 0.12746 | ser-phe | 0.115 |
| cys-tyr | 0.401 | gly-ile | 0.405 | asn-tyr | 0.12025 | val-val | 0.107 |
| ala-gln | 0.383 | pro-leu | 0.392 | ala-leu | 0.10727 | cys-lys | 0.098 |
| asn-phe | 0.376 | lys-phe | 0.392 | pro-ser | 0.10662 | pro-gln | 0.096 |
| asn-leu | 0.373 | gln-ser | 0.387 | cys-ser | 0.10641 | asn-his | 0.094 |
| arg-tyr | 0.369 | val-val | 0.385 | ser-glu | 0.09534 | cys-gly | 0.094 |
| lys-ile | 0.368 | ser-phe | 0.379 | gln-ser | 0.09214 | thr-thr | 0.093 |
| ala-phe | 0.368 | asn-trp | 0.375 | glu-ile | 0.09055 | gln-his | 0.091 |
| ala-asp | 0.367 | asp-val | 0.375 | thr-his | 0.09034 | pro-lys | 0.088 |
| arg-glu | 0.365 | asp-leu | 0.374 | thr-asp | 0.08029 | ala-phe | 0.087 |
| asn-arg | 0.364 | ser-leu | 0.37 | thr-asn | 0.06146 | thr-leu | 0.081 |
| lys-glu | 0.364 | thr-ile | 0.367 | glu-leu | 0.05922 | ala-lys | 0.08 |
| leu-ile | 0.354 | val-ile | 0.364 | trp-trp | 0.05407 | asn-arg | 0.079 |
| tyr-asp | 0.35 | glu-leu | 0.364 | asn-phe | 0.05406 | ser-asp | 0.075 |
| gln-tyr | 0.348 | ser-arg | 0.362 | pro-glu | 0.04856 | gln-glu | 0.074 |
| met-arg | 0.343 | lys-leu | 0.362 | ser-ser | 0.04606 | cys-gln | 0.065 |
| ser-val | 0.342 | thr-val | 0.361 | arg-leu | 0.04463 | thr-gln | 0.065 |
| met-his | 0.34 | arg-asp | 0.359 | arg-asp | 0.04427 | phe-asp | 0.064 |
| met-gln | 0.335 | ser-ile | 0.359 | tyr-leu | 0.04037 | glu-leu | 0.06 |
| arg-ile | 0.32 | lys-gly | 0.356 | tyr-ile | 0.03939 | cys-thr | 0.056 |
| val-gly | 0.314 | val-leu | 0.355 | glu-asp | 0.03712 | lys-glu | 0.054 |
| leu-gly | 0.309 | gln-trp | 0.355 | his-his | 0.0352 | tyr-leu | 0.048 |
| met-ile | 0.308 | arg-leu | 0.354 | ser-arg | 0.03249 | val-ile | 0.048 |
| cys-lys | 0.308 | pro-trp | 0.35 | cys-phe | 0.02772 | cys-ile | 0.042 |
| ser-glu | 0.306 | met-asp | 0.35 | val-leu | 0.02515 | glu-ile | 0.039 |

| Res-Res | CC | Res-Res | CC | Res-Res | CC | Res-Res | CC |
|---|---|---|---|---|---|---|---|
| arg-asp | 0.3 | trp-trp | 0.348 | gln-asp | 0.01386 | ser-glu | 0.036 |
| cys-asp | 0.299 | ala-phe | 0.345 | his-trp | 0.01223 | asn-ile | 0.032 |
| ala-thr | 0.299 | cys-trp | 0.343 | cys-thr | 0.01204 | cys-glu | 0.031 |
| ala-leu | 0.297 | ser-gly | 0.342 | asn-asp | 0.01124 | lys-tyr | 0.029 |
| glu-ile | 0.296 | thr-asn | 0.341 | pro-phe | 0.01087 | his-asp | 0.026 |
| asn-his | 0.295 | lys-asp | 0.337 | val-gly | 0.01045 | pro-asp | 0.024 |
| ile-ile | 0.295 | thr-phe | 0.333 | asn-arg | 0.00892 | lys-leu | 0.02 |
| thr-gln | 0.294 | lys-ile | 0.329 | his-leu | 0.00656 | cys-asp | 0.019 |
| met-asn | 0.293 | trp-asp | 0.327 | arg-arg | 0.00293 | lys-val | 0.019 |
| cys-val | 0.277 | asp-ile | 0.327 | gln-gln | -0.0002 | asp-ile | 0.013 |
| cys-leu | 0.276 | cys-val | 0.325 | phe-asp | -0.0004 | thr-asn | 0.01 |

**CC: Correlation Coefficient.**

**Table-3 Legend:** Enlisted above are 50 residue-residue interactions, who, along the evolutionary timeline have recorded maximum increment in the extent of their symmetry in structural class-specific manner. Any pairwise interaction that registers among these enlisted 50 (and therefore, counted among evolutionarily most-favored) in more than one structural class, is termed as an 'acceptable evolutionary tunnel'. Only one instance of acceptable evolutionary tunnel is found that is favored in all four structural classes, it is ARG-ASP symmetry. Acceptable evolutionary tunnels those are favored in three structural classes are (total 13): VAL-GLY, ALA-LEU, LYS-ILE, THR-VAL, ALA-PHE, SER-GLU, GLU-ILE, ASN-ARG, ASP-LEU, LYS-LEU, PHE-ASP, THR-ASN, GLU-LEU. Acceptable structural tunnels those are favored in two structural classes are (total 38): HIS-HIS, LYS-VAL, CYS-SER, ALA-LYS, TYR-LEU, CYS-THR, ALA-GLN, PRO-TYR, ASN-PHE, THR-PHE, GLN-LEU, LYS-ASP, PRO-PHE, ASN-TRP, GLN-SER, TRP-TRP, SER-ARG, ARG-LEU, VAL-LEU, GLN-ARG, ASN-ILE, LEU-GLY, CYS-LYS, ASN-HIS, LYS-GLU, THR-GLN, CYS-ASP, ALA-VAL, SER-TRP, THR-LEU, SER-PHE, VAL-VAL, VAL-ILE, ASP-ILE, SER-VAL, GLN-TYR, ARG-GLU, CYS-VAL.

### Table-4: Non-Acceptable Evolutionary Tunnels

| ALL-ALPHA | | ALL-BETA | | ALPHA/BETA | | ALPHA+BETA | |
|---|---|---|---|---|---|---|---|
| **Res-Res** | **CC** | **Res-Res** | **CC** | **Res-Res** | **CC** | **Res-Res** | **CC** |
| asp-val | 0.08867 | cys-gly | 0.1435 | gln-trp | -0.303 | glu-glu | -0.19182 |
| thr-pro | 0.08575 | gln-lys | 0.1419 | pro-asp | -0.309 | met-tyr | -0.19344 |
| asn-gly | 0.08468 | ala-pro | 0.1377 | tyr-gly | -0.312 | ala-arg | -0.1961 |
| ser-trp | 0.08465 | arg-arg | 0.1373 | asn-gly | -0.313 | met-lys | -0.19873 |
| his-ile | 0.08455 | lys-tyr | 0.1349 | pro-gln | -0.315 | ser-arg | -0.20389 |
| tyr-tyr | 0.08404 | phe-ile | 0.1345 | cys-pro | -0.316 | gln-trp | -0.2051 |
| ser-gly | 0.08355 | ala-arg | 0.1328 | asp-val | -0.332 | tyr-gly | -0.20634 |
| pro-lys | 0.08141 | tyr-leu | 0.1306 | met-ile | -0.334 | ala-asp | -0.20697 |
| lys-tyr | 0.07933 | pro-val | 0.1273 | trp-asp | -0.34 | thr-lys | -0.20847 |
| tyr-trp | 0.07813 | his-val | 0.1209 | his-gly | -0.343 | trp-ile | -0.21053 |
| pro-asn | 0.0727 | met-trp | 0.1197 | cys-glu | -0.346 | ser-ile | -0.21212 |
| thr-asp | 0.06749 | gln-asn | 0.119 | phe-trp | -0.346 | ala-thr | -0.21513 |
| trp-leu | 0.06732 | thr-pro | 0.1158 | lys-lys | -0.353 | trp-trp | -0.21758 |
| glu-glu | 0.06397 | thr-his | 0.1142 | ser-phe | -0.357 | thr-tyr | -0.21979 |
| cys-trp | 0.06241 | pro-lys | 0.1086 | ala-met | -0.361 | phe-tyr | -0.22631 |
| gln-trp | 0.0572 | gln-arg | 0.1071 | glu-glu | -0.368 | phe-val | -0.22836 |
| ser-tyr | 0.05614 | phe-val | 0.106 | met-thr | -0.37 | leu-ile | -0.22906 |
| asn-ser | 0.05555 | arg-trp | 0.1058 | gln-tyr | -0.373 | lys-ile | -0.23085 |
| pro-gln | 0.04918 | met-glu | 0.1033 | glu-val | -0.38 | ala-tyr | -0.2342 |
| lys-leu | 0.04477 | pro-arg | 0.1026 | phe-glu | -0.381 | met-met | -0.23956 |
| glu-leu | 0.03779 | met-ser | 0.1022 | met-his | -0.383 | leu-leu | -0.24014 |
| asn-lys | 0.03299 | met-arg | 0.0994 | ala-pro | -0.383 | ser-leu | -0.24064 |
| ser-ile | 0.02633 | trp-val | 0.0986 | met-phe | -0.383 | tyr-trp | -0.24278 |
| asp-asp | 0.02442 | glu-asp | 0.0966 | his-tyr | -0.397 | asp-asp | -0.24439 |
| met-phe | 0.0212 | cys-glu | 0.087 | ala-arg | -0.412 | his-trp | -0.24648 |
| ser-arg | 0.01794 | trp-ile | 0.0848 | his-glu | -0.42 | pro-trp | -0.25128 |
| asn-asp | 0.01119 | cys-thr | 0.076 | cys-lys | -0.42 | ala-his | -0.25802 |
| his-trp | 0.00891 | cys-met | 0.0721 | phe-ile | -0.423 | met-gly | -0.25848 |
| cys-met | 0.00653 | met-asn | 0.06 | ile-ile | -0.425 | cys-leu | -0.26159 |
| met-met | 0.00387 | met-ile | 0.0571 | tyr-glu | -0.462 | arg-his | -0.26245 |
| asn-trp | -0.0026 | met-phe | 0.055 | leu-leu | -0.462 | thr-trp | -0.26271 |
| his-asp | -0.0075 | ser-lys | 0.0498 | cys-asp | -0.464 | thr-phe | -0.26655 |
| gln-lys | -0.0078 | ala-his | 0.0445 | met-trp | -0.465 | his-leu | -0.27956 |
| phe-val | -0.0097 | pro-pro | 0.0304 | trp-val | -0.471 | thr-arg | -0.29473 |
| cys-pro | -0.0111 | his-lys | 0.0229 | met-val | -0.489 | his-ile | -0.29547 |
| thr-trp | -0.0146 | met-gly | 0.0177 | cys-val | -0.491 | val-leu | -0.30908 |
| trp-trp | -0.018 | asn-his | 0.0171 | trp-ile | -0.492 | ala-met | -0.3159 |
| asp-gly | -0.0203 | phe-phe | 0.0122 | tyr-trp | -0.496 | glu-asp | -0.3192 |
| ser-his | -0.0451 | gln-his | 0.0111 | asn-asn | -0.531 | pro-pro | -0.33046 |
| cys-his | -0.0547 | cys-phe | 0.0015 | phe-tyr | -0.531 | ser-val | -0.33973 |
| pro-arg | -0.0563 | arg-lys | -0.0112 | met-met | -0.532 | pro-his | -0.34834 |
| his-gly | -0.0598 | pro-his | -0.0156 | phe-phe | -0.547 | met-arg | -0.36147 |
| gln-his | -0.068 | cys-asp | -0.0279 | tyr-tyr | -0.562 | tyr-tyr | -0.36742 |
| cys-cys | -0.0817 | cys-lys | -0.0595 | ala-gly | -0.563 | ala-pro | -0.36787 |
| pro-val | -0.0928 | cys-gln | -0.0808 | pro-val | -0.565 | met-asn | -0.3836 |
| met-trp | -0.1119 | cys-arg | -0.0828 | ala-trp | -0.6 | his-val | -0.39041 |
| trp-gly | -0.1781 | cys-his | -0.0927 | ala-phe | -0.616 | lys-phe | -0.40413 |
| thr-his | -0.1898 | met-met | -0.1145 | pro-pro | -0.64 | pro-phe | -0.42658 |
| phe-phe | -0.1921 | met-his | -0.1878 | arg-val | -0.658 | met-glu | -0.45917 |
| pro-trp | -0.3265 | cys-pro | -0.2002 | ala-ala | -0.678 | met-phe | -0.5619 |

**CC:** Correlation Coefficient.

**Table-4 Legend:** Enlisted above are 50 residue-residue interactions, who, along the evolutionary timeline have recorded maximum decrement in the extent of their symmetry in structural class-specific manner. Any pairwise interaction that registers among these enlisted 50 (and therefore, counted among evolutionarily least-favored) in more than one structural class, is termed as a 'non-acceptable evolutionary tunnel'. Two non-acceptable evolutionary tunnels that are universally disliked by evolutionary constraints in all four structural classes are found to be: MET-MET and MET-PHE. Non-acceptable evolutionary tunnels those are disliked in three structural classes are (total 13): GLN-TRP, GLU-GLU, TYR-TYR, TYR-TRP, ALA-PRO, ALA-ARG, TRP-ILE, PRO-PRO, CYS-PRO, PRO-VAL, MET-TRP, PHE-PHE, PHE-VAL. Acceptable structural tunnels those are favored in two structural classes are (total 40): TYR-GLY, ALA-MET, PHE-TYR, LEU-LEU, ASP-VAL, ASN-GLY, PRO-GLN, HIS-GLY, PHE-ILE, MET-ILE, CYS-GLU, MET-HIS, TRP-VAL, CYS-

LYS, CYS-ASP, HIS-ILE, SER-ARG, SER-ILE, TRP-TRP, ASP-ASP, HIS-TRP, PRO-TRP, THR-TRP, HIS-VAL, MET-GLU, MET-ARG, GLU-ASP, ALA-HIS, MET-GLY, MET-ASN, PRO-HIS, THR-PRO, GLN-LYS, LYS-TYR, PRO-LYS, THR-HIS, PRO-ARG, CYS-MET, GLN-HIS, CYS-HIS.

### *3.3: Uniformity of relative contributions of certain STs and ETs*

How expected is it to find that a structural or evolutionary tunnel is favored or disfavored by *equivalent relative extents* in two or as many as three structural classes? Since geometrical constraints on proteins belonging to any structural class characterize the uniqueness of that structural class's scheme to ensure mechanical and energetic stability, observing a pairwise interaction symmetry to be favored or disfavored by different sets of structural constraints in equivalent relative extents – seems a highly unlikely proposition. However, it has been found that there are as many as 5 cases where STs are favored by two or three structural classes by identical relative extent. For example, ARG-LYS and ASP-LEU symmetries are found to be liked by all-$\alpha$, all-$\beta$, $\alpha/\beta$ and all-$\beta$, $\alpha/\beta$, $\alpha+\beta$ respectively – in equal degree; while the former features as the most-favored ST in 3 structural classes, the later features as the 3$^{rd}$ most-favored ST in 3 structural classes. Similar occurrences can be found for GLU-ASP (2$^{nd}$ most-favored in all-$\alpha$, $\alpha+\beta$), PHE-GLU (4$^{th}$ most-favored in all-$\alpha$, $\alpha/\beta$) and ARG-PHE (10$^{th}$ most-favored in all-$\beta$, $\alpha/\beta$). The exact contribution of pairwise interaction symmetries to structural stability of respective classes are found to be different, but in spite of difference in magnitude of contributions, the degree to which the aforementioned tunnels are favored across structural classes – are found to be same.

More such equivalent cases are found in nature's way of disapproving certain STs. A total of 9 cases could be detected amongst non-acceptable STs. Each of these 9 STs is disliked in two structural classes by identical relative degree. Amon these, the case of MET-TRP symmetry stands out most prominently, it features both as 4$^{th}$ least-favored non-acceptable ST in $\alpha/\beta$ and $\alpha+\beta$ class of proteins, and also as 6$^{th}$ least-favored ST in all-$\alpha$ and all-$\beta$ structural classes. TRP-TRP interaction symmetry is detected as the least favored interaction symmetry in all-$\alpha$ and $\alpha+\beta$ class of proteins. Barring these, MET-MET (6$^{th}$ least-favored ST across all-$\beta$ and $\alpha/\beta$), HIS-TRP (7$^{th}$ least-favored ST in all-$\alpha$ and all-$\beta$), ILE-ILE (9$^{th}$ least-favored ST in $\alpha/\beta$ and $\alpha+\beta$), CYS-TYR (10$^{th}$ least-favored ST in all-$\alpha$, $\alpha/\beta$), LEU-LEU (14$^{th}$ least-favored ST in all-$\beta$, $\alpha/\beta$) and LEU-ILE (15$^{th}$ least-favored ST in all-$\beta$, $\alpha/\beta$) – could be detected.

*Remarkably, all the STs that are either favored or disfavored in equal relative extent are found to be clustered within either the most-favored 10 tunnels or within the least-favored 15 tunnels. – This indicates a clear-cut strategy of nature to treat certain pairwise interaction symmetries as equally favored and certain others as equally disfavored across structural classes, regardless of the specific constraints that mark any of the structural classes.*

Occurrences of ETs registering equal relative extents in being liked or disliked by structural classes are found to be less common. But then, as one may note easily, favoring or disfavoring the ETs with equal relative extent is an (almost) impossible occurrence to expect at the first place; because, unlike the straightforward ST analysis, the ET analysis attempted to detect the cases where pairwise residual interaction symmetry in proteins belonging to evolutionarily sorted 76 SCOP-folds have increased or decreased in exactly the same relative margin for different SCOP-classes. Astonishingly, such (incredible) equivalences are found in as many as 5 (2 favored, 3 disfavored) cases. Amongst the non-acceptable ETs, PRO-VAL interaction symmetry is found to be the 6$^{th}$ least favored ET across all-$\alpha$ and $\alpha/\beta$ proteins, TYR-TYR is found to be the 8$^{th}$ least favored in both $\alpha/\beta$ and $\alpha+\beta$, ASP-ASP is detected to be the 27$^{th}$ least favored ET in both all-$\alpha$ and $\alpha+\beta$. Amongst the acceptable ETs, ARG-ASP is found to be equally (32$^{nd}$) favored across all-$\beta$ and $\alpha/\beta$ class of proteins, ASP-ILE is found to be equally (49$^{th}$) favored in all-$\alpha$ and $\alpha+\beta$ proteins.

***3.4. Meeting ground between Structural Tunnels and Evolutionary Tunnels:*** There are merely 25 acceptable STs and 50 acceptable ETs. On the other end of the spectrum, there are only 25 non-acceptable STs and 50 non-acceptable ETs. Thus, the total number of acceptable and non-acceptable tunnels, in each case, adds up to only 75. – This is an exceptionally small subset of the total number of enormous possibilities, given especially the combinatorial scope of matching between $\binom{210}{25}C)^4$ STs and $\binom{210}{50}C)^4$ ETs. *Thus, the probability that, certain acceptable STs will feature among the set of acceptable ETs and certain non-acceptable STs will feature amongst the set of non-acceptable ETs, qualifies as one truly rare, and therefore, truly significant.* Such common elements among acceptable tunnels represent the pairwise residual interaction symmetries that not only prevail across different structural/geometric constraints, but also as ones that satisfy the diverse set of criteria to efficiently fit the structural class-specific evolutionary landscape. Conversely, common elements amongst non-acceptable tunnels represent the pairwise residual interaction symmetries that are universally not-favored by geometries of structural classes and are universally less efficient to fit into evolutionary landscape.

The MET-MET interaction symmetry is found to be the only one that is disliked truly universally (viz., non-acceptable #ST = non-acceptable #ET = 4) both structurally and evolutionarily (**Table-2** and **Table-4**). In contrast, no pairwise interaction symmetry could be detected as being truly universally liked. ASP-LEU interaction symmetry is found to be the closest to being universally liked, characterized by acceptable #ST= 4 and acceptable #ET=3. Symmetries of SER-GLU, GLU-ILE, PHE-ASP interactions were detected to be conducive, since each one of them registered acceptable #ST=3 and acceptable #ET=3. In contrast, MET-TRP, PHE-PHE and CYS-PRO symmetries were found to be highly disliked both by structural as well as evolutionary constraints; as a result, while MET-TRP and PHE-PHE were detected to have non-acceptable #ST=4 and non-acceptable #ET= 3, CYS-PRO symmetry recorded non-acceptable #ST=non-acceptable #ET= 3. Interestingly, while a (somewhat) larger population of residue interaction symmetry (LEU-LEU, TRP-TRP, HIS-TRP, CYS-MET, CYS-HIS) is found to populate the slot non-acceptable #ST=4 and non-acceptable #ET=2, no symmetry could be detected to populate this slot for acceptable tunnels. On the other hand, though no interaction symmetry categorized itself as non-acceptable #ST=3 and non-acceptable #ET=2, ASP-ILE symmetry registered acceptable #ST = 3 and acceptable #ET = 2. **Table-5** contains further details in this line.

Certain residual interaction symmetries were conspicuous in their absence from meeting ground between ETs and STs. For example, though ARG-ASP was noted to be the most significant acceptable ET with #ET=4, it was not found among the acceptable STs. A large set of residual interaction symmetries (that of VAL-GLY, ALA-LEU, THR-VAL, ALA-PHE, ASN-ARG, LYS-LEU and THR-ASN) were accommodated under varied constraints of evolutionary fitness, registering a non-trivial #ET = 3. But none of these was found among the ST population. These tend to imply that while nature is quite decisive in favoring or not-favoring certain residual interaction symmetries, the process of selection of interaction symmetries that are compatible for both evolutionary and structural constraints is perhaps still continuing.

To explore the inter-relationship between ET and ST further, the possible existence of mapping scheme between the pattern of differences in structural class-specific magnitudes of STs and ETs was probed too (**Suppl. Mat.-2**). However, no such general mapping could be found to relate the magnitude of STs to the magnitude of ETs.

**Table-5: Meeting ground between (acceptable STs and ETs) and (non-acceptable STs and ETs):**

|  | Acceptable ST (tunnel strength=4) | Acceptable ST (tunnel strength=3) | Acceptable ST (tunnel strength=2) |
|---|---|---|---|
| **Acceptable ET (tunnel strength=4)** | No Such Case. | No Such Case. | No Such Case. |
| **Acceptable ET (tunnel strength=3)** | ASP-LEU | SER-GLU, GLU-ILE, PHE-ASP | LYS-ILE, GLU-ILE |
| **Acceptable ET (tunnel strength=2)** | No Such Case. | ASP-ILE | LYS-VAL, ASN-ILE |
|  | **Non-Acceptable ST (tunnel strength=4)** | **Non-Acceptable ST (tunnel strength=3)** | **Non-Acceptable ST (tunnel strength=2)** |
| **Non-Acceptable ET (tunnel strength=4)** | MET-MET | No Such Case. | MET-PHE |
| **Non-Acceptable ET (tunnel strength=3)** | MET-TRP, PHE-PHE | CYS-PRO | TYR-TRP, TRP-ILE |
| **Non-Acceptable ET (tunnel strength=2)** | LEU-LEU, TRP-TRP, HIS-TRP, CYS-MET, CYS-HIS | No Such Case. | PHE-ILE |

*3.5. Of Tunnels and Chou-Fasman(CF) propensities:* Amino acids and secondary structures have mutual liking and disliking for each other (Chou and Fasman, 1974). Results (*section 3.1-3.4*) establish that STs and ETs serve as common platforms for protein organization, connecting structural and evolutionary checkpoints. Restricting ourselves only to pairwise-symmetries in all-$\alpha$ and all-$\beta$ classes, one may ask: do the components (viz. individual residues) of acceptable STs and ETs in all-$\alpha$ and all-$\beta$ class, demonstrate high degree of correspondence with their CF propensities for $\alpha$-helices and $\beta$-sheets? – If they do, it will imply that nature's *modus operandi* to favor some of the residual interaction symmetries (in the form of acceptable STs and ETs) stem from the level of single residues and continue all the way to the functioning state of quaternary-structure of the protein. But in case if the aforementioned correspondence is not observed, it will imply that while the CF propensities basically enlist the stereochemical preferences of single amino acids (like, the size of the side-chain, etc.), nature's organizational principles, both in structural and evolutionary terms, assign more onus to the *interactions* amongst the residues rather than on the individual residues themselves.

To identify the correct one of these two possibilities, populations of residues in acceptable STs and ETs in all-$\alpha$ and all-$\beta$ class were estimated at first. Then, correlation coefficient was calculated between, the all-$\alpha$ (or all-$\beta$) class of tunnel population of individual residues and CF propensities of individual residues in $\alpha$-helix (or $\beta$-sheet). Obtained results ($Correl.Coeff_{Acceptable-ST}^{All-\alpha\ and\ \alpha-helix\ CF} = 0.227$, $Correl.Coeff_{Acceptable-ST}^{All-\beta\ and\ \beta-sheet\ CF} = -0.192$, $Correl.Coeff_{Acceptable-ET}^{All-\alpha\ and\ \alpha-helix\ CF} = 0.245$, $Correl.Coeff_{Acceptable-ET}^{All-\beta\ and\ \beta-sheet\ CF} = 0.349$) revealed that there's very little dependence between the CF propensity of a residue in secondary structures and the number of occurrences of this residue as part of either ST or ET. Details can be found in **Suppl. Mat.-3**.

The fact that most symmetric residual interactions in all-$\alpha$ or all-$\beta$ class do not (in general) involve residues with high CF propensity for $\alpha$-helix or $\beta$-sheet, tends to suggest that nature's methods to optimize the geometrical features of quaternary structures and also its methods to ensure effective functionality of the native protein to suit evolutionary fitness landscape – starts not at the level of individual residues but at the level of interactions amongst residues. Concrete evidence to such assertion is found from the observations that VAL, known for its superlative sheet-forming tendencies, occurs only in 6% of acceptable all-$\beta$ STs; ALA, known for its high helix-propensity, occurs only in 4% of acceptable STs and in 7% of acceptable ETs in all-$\alpha$ class; similarly MET, known for its equally strong helix-forming credentials, occurs not even once in the acceptable

all-$\alpha$ STs and only in 9% of acceptable ETs in all-$\alpha$ class. In stark contrast, PRO, never known for its helix-loving credentials, is found to occur in 10% of acceptable STs and 6% of acceptable ETs in all-$\alpha$ class; similarly, ASP, seldom considered for its sheet-forming propensity, occurs in 10% of acceptable STs and 8% of acceptable ETs in all-$\beta$ class.

However, no easy pattern can be found from these observations; because unlike the case of ALA and MET, GLU (known for strong helix-forming potential) is found to be a part of 16% of acceptable all-$\alpha$ STs. But then again, though LYS and TRP are known to have similar CF propensity for helix, LYS occurs in as many as 18% of acceptable STs in all-$\alpha$ class, whereas TRP occurs absolutely nowhere in helically acceptable STs. – All of these signify clearly that stereochemical properties of an individual residue are either insufficient or totally irrelevant when used for investigating the symmetry in pairwise interactions that this residue may be part of, especially when investigating whether such a pairwise interaction is favored or disfavored by structural or evolutionary constraints at the level of the functional state of the protein.

***3.6. To what extent can we detect commonality among tunnels by studying residual properties? In other words, to what extent STs and ETs demonstrate strong emergence?*** It is known that most local interactions among residues are mediated by direct contact between residues (or by water molecules or hydration shells). Dispersion forces between different atom types are known to have their potential wells predominantly in the range 2.5-5.0 Å, with very weak interactions at distances larger than 5.0-7.0 Å (Nemethy et al., 1983). The limit for hydrogen bonds is reported to be ~ 4.2 Å (Stickle et al., 1992). Coulombic interactions are known to be screened due to the high dielectric constant of the solvent but may still have a long range ($\leq$15 Å). – Given this level of complexity of the interactions themselves, it becomes difficult to decipher the rules of symmetry-promotion and symmetry-preservation in residual interactions by investigating the properties of individual residues alone. Nevertheless an attempt was made in this work to investigate two related questions. First, why, pairwise interactions with high extent of symmetry are produced only by a small subset of 20 residues (- so that ST and ET compositions consistently demonstrate bias for certain residues, in contrast to possible homogeneity of 20 populations as tunnel-components)? Second, to what extent, the properties of STs and ETs cannot be traced back to properties of individual residues? (viz., to what extent STs and ETs are strong-emergent features?)

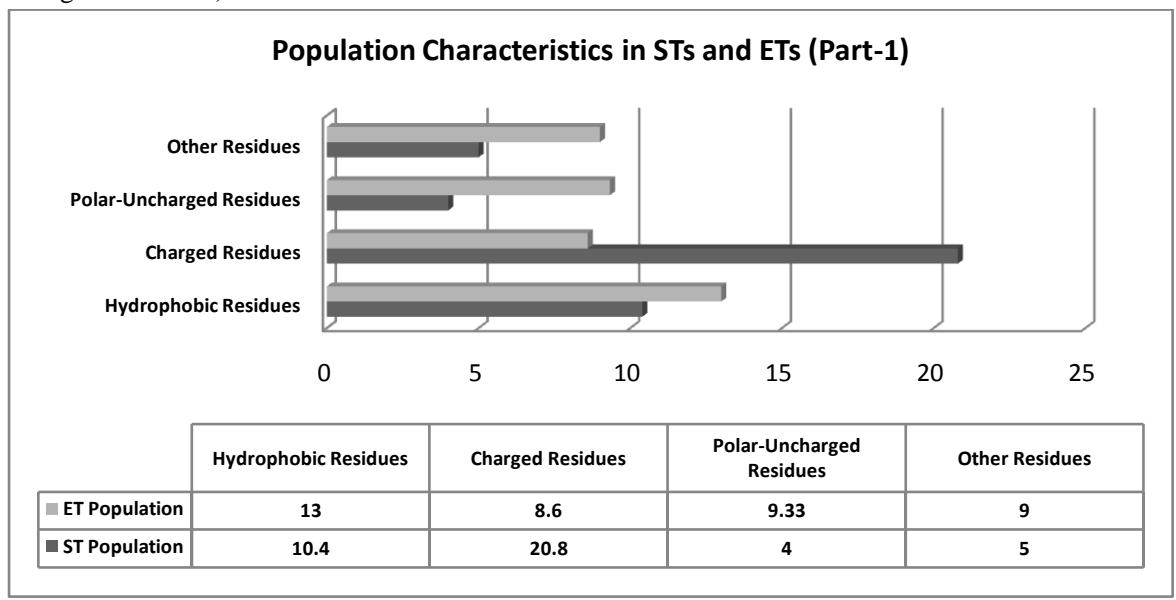

**Population Characteristics in STs and ETs (Part-1)**

|  | Hydrophobic Residues | Charged Residues | Polar-Uncharged Residues | Other Residues |
|---|---|---|---|---|
| ET Population | 13 | 8.6 | 9.33 | 9 |
| ST Population | 10.4 | 20.8 | 4 | 5 |

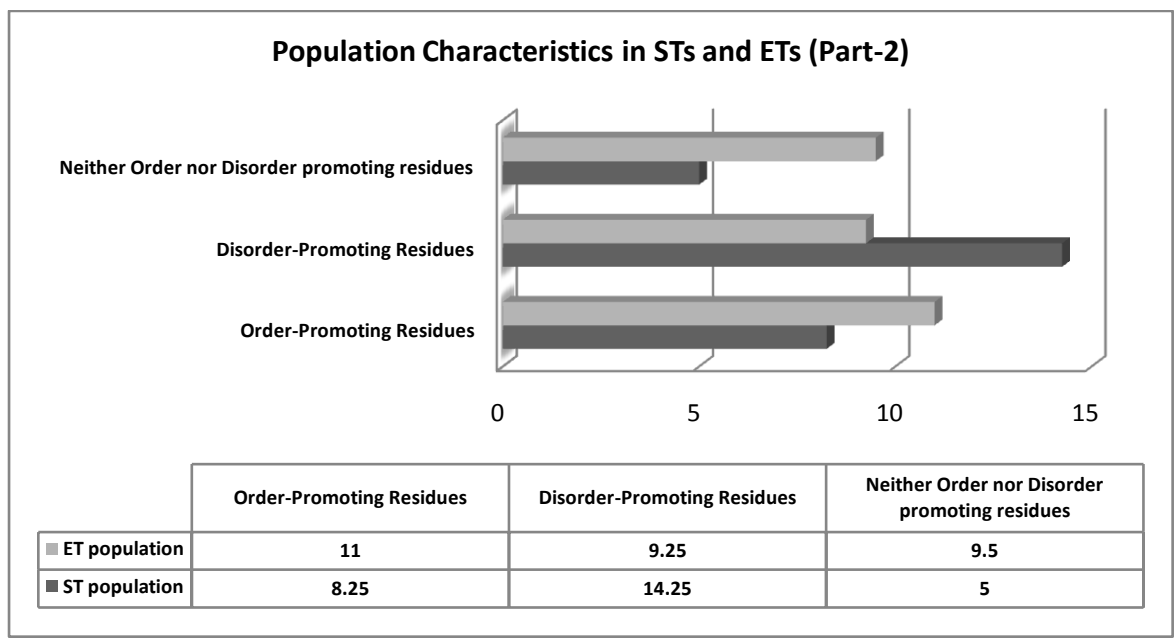

**Figure-1: Comparative average population of residues in each type.**

**Figure-1 Legend:** Average occurrence of residues in each class of residues in plotted here. Details about the types of residues are provided in the main text.

*3.6.1. Population characteristics of charged, hydrophobic, polar-uncharged residues in STs and ETs:* **Fig-1** presents the distribution of comparative percentage population of different types of residues; we study not only the population percentage of hydrophobic, charged and polar uncharged residues, but also the population of residues that are either order-promoting or disorder-promoting (Dunker et al., 2001). **Fig-1.(Part-1)** describes a distinctly favoring bias for charged residues as the component of STs. It depicts that on an average, each of the charged residues (ASP, GLU, ARG, LYS, HIS) constitutes ~21% of ST components. In comparison, each hydrophobic residue (VAL, LEU, ILE, MET, PHE) constitutes 10.4% of ST components; each polar-uncharged residue (CYS, SER, THR, GLN, ASN, TYR) constitutes only 4% of ST components and other residues (PRO, GLY, ALA, TRP), on an average, constitutes 5% of ST components. Closer inspection of **Table-1-to-Table-4** data reveals that preference for charged residues in ST components assumes more noticeable proportion if HIS contribution is not considered; whereby, the average contribution of other charged residues (ASP, GLU, ARG and LYS) to ST components becomes 25%. Similarly, average contribution of hydrophobic residues to ST records 13%, if the case of MET (which contributes 0% to both all-$\alpha$ and all-$\beta$ STs) is ignored. Interestingly, unlike the case of STs, no preference for charged residues could be found in ET population; the remarkably unbiased and comparable nature of residue population in ETs is demonstrated in **Fig-1.(Part-1)**. *Why did nature consider the symmetry in interactions amongst 4 charged residues to be so important for structural considerations and yet, why for evolutionary optimization, did nature consider the interactional symmetry among hydrophobic residues as importantly as that amongst charged residues (and (separately) amongst polar-uncharged residues) – is difficult to ascertain from the obtained data.*

*3.6.2: Population characteristics of ST and ET residues that are order-promoting, disorder-promoting, or are neither order-promoting nor disorder-promoting:* From a completely different perspective, degree of symmetry in interactions among 'order-promoting' and 'disorder-promoting' residues were studied **Fig-1.(Part-2)**. While the average contribution of each of the disorder-promoting residues (ALA, ARG, GLY,

GLN, SER, PRO, GLU, LYS) to acceptable-ST composition is found to be 14.25%, each of the order-promoting residues (TRP, CYS, PHE, ILE, TYR, VAL, LEU, ASN) is found to have contributed only 8.25% to acceptable-ST compositions. Interestingly, the residues that promote neither order nor disorder (viz., HIS, MET, THR, ASP), are found to contribute a trifle 5% per-residue to acceptable-ST composition. However, in striking similarity to *Section-3.6.1*-results, populations of acceptable-ET-residues that promote order, disorder and neither of these two – are found to be comparable and very similar. While every order-promoting residue on an average is found to constitute 11% of ET, a disorder-promoting residue constitutes 9.25% and each residue that promotes neither order nor disorder is found to constitute 9.5%. Though one may naively point out the occurrence of some hydrophobic and charged residues in the lists of 'order-promoting' and 'disorder-promoting' residues respectively, such a mapping between two classes of residues is far from being trustworthy. The presence of small but definite favoring bias for 'disorder-promoting' residues (over the 'order-promoting' ones) among the acceptable-ST components, therefore, strikes as a noticeable pattern. However, just like the case of *Section-3.6.1*, why does nature favor all three classes of residues in (almost) similar extent in acceptable-ET – could not be answered from the present set of data.

### *3.6.3: Does relative abundance of residues influence selection of tunnel residues?*

It is not illogical to probe if the global stoichiometry of amino acids influences nature's decision to accommodate or exclude certain residues in STs and ETs. Table of residual stoichiometry describes the relative abundance of these residues in biological universe. Could it be the case that more abundant residues are chosen as components of acceptable-STs and acceptable-ETs? Detection of similar trends in stoichiometric and tunneling populations would have suggested a mechanistic basis for abundance of certain residues over others; in contrast, detection of different trends would have implied that nature's preference for certain residues as component of acceptable-STs and acceptable-ETs is (probably) not dependent upon the bulk at which they are produced, but are (probably) more dependent on the ability of these residues to participate in interactions that promote and preserve symmetry.

Results, however, did not present a clear signal. Correlation coefficient between the global stoichiometric proportions of residues (Voet and Voet, 2010) and residue population in all-$\alpha$ acceptable-ST was found to be 0.31; that between the former and residue population in all-$\beta$ acceptable-ST was found to be 0.33. Though highly similar, based on the magnitudes of correlation coefficients, no clear inference on the nature of dependency between tunnel residues and the pool of more abundant residues could be derived. Interestingly, while correlation between stoichiometric proportions of residues and populations of residues in all-$\alpha$ acceptable-ET and is found to be absolutely non-existent (correlation coefficient =0.02), that between the former and acceptable-ET population of all-$\beta$ residues is noted to be 0.48. This implied that while the selection of all-$\alpha$ residues suitable to be accommodated in acceptable-ET is not at all dependent upon relative availability of residues, selection criterion for all-$\beta$ residues as components of acceptable-ET is, to a non-trivial extent, dependent on the relative availability of residues. However, no satisfactory explanation behind nature's implementing two different strategies for managing evolutionary constraints in as many structural classes could be found from the present set of data.

### *3.6.4. Fragments of patterns in residual interactions that feature in STs and ETs:*

Barring the (somewhat) definite preference for the charged residues in STs, no general trend was found to describe the interplay between tunnels and properties of individual amino acids. Indeed, the absence of any simplistic (stereotype) pattern in both ST and ET composition can easily be identified as a salient feature of tunnels. In order to demonstrate the strong emergent nature of tunnels, we chose to proceed by depicting the inherent inadequacy of attempts to understand tunnels with 'residue-to-secondary structure' mapping construct, in few cases. Barring few exceptions ((Banerji and Ghosh, 2010) and references therein) studies on emergent features in protein structural organization isn't common. The present work demonstrates why symmetry in pairwise-interactions can be considered as an emergent feature at the level of quaternary structures.

### *3.6.4.1. (Near)-absence of HIS from all-α and all-β STs and ETs:*

We start by discussing the prominent (near-complete) absence of HIS from STs and ETs of both all-$\alpha$ and all-$\beta$ structures. HIS occurs in only 4% of all-$\alpha$ ST components and it does not feature in all-$\beta$ STs at all; it contributes to all-$\alpha$ and all-$\beta$ ET components by 3% and 0% respectively. In comparison, the %-population of ARG in all-$\alpha$ and all-$\beta$ STs are found out to be 12% and 10%, in all-$\alpha$ and all-$\beta$ ETs ARG occurred in 9% and 4% of the cases. LYS occurred in all-$\alpha$ and all-$\beta$ STs in 18% and 14% of cases and in all-$\alpha$ and all-$\beta$ ETs in 4% and 5% of the cases. HIS's CF-propensity for helix and sheet is comparable, 1.00 and 0.87 respectively; and is equally unremarkable as the CF-propensity for helix and sheet of ARG (0.98 and 0.93) and LYS (1.14 and 0.74) respectively. Yet, ARG and LYS is found to occur heavily in all-$\alpha$ and all-$\beta$ STs and ETs, but not HIS. We note that HIS is unique because it contains a positively charged imidazole functional group. Since the unprotonated imidazole is nucleophilic, it serves as a general base, but the protonated form serves as a general acid. Due to this useful dual role, HIS is known to stabilize the folded structures and occurs as a common participant in enzyme catalyzed reactions. Strangely, HIS is found to feature rarely in all-$\alpha$ and all-$\beta$ STs and ETs. – Such minimal representation suggests that, as a property, symmetry in pairwise interaction cannot be mapped on to the usefulness of a single residue, nor can it be explained with CF propensity. The disproportionately small representation of HIS also underlines that though one finds a fragment of pattern in high population of charged residues in STs; such 'charged residue-to-tunnel' mapping is not general.

### *3.6.4.2. Does tunneling population depend on size of amino acids?*

To what extent, if at all, accommodation of a residue in STs and ETs depends on its size? In other words, does the symmetry of residual interaction depend on the size of residues? – We do note that across STs in all four SCOP classes, TRP and GLY, two extremes, feature minimally (0 and 1, respectively) among the 25 most-favored STs. However, while STs containing TRPs occur as many as 8 times as least-favored 25 STs, GLY doesn't feature at all in any of the least-favored STs, leaving no pattern to relate residue-size to residual-interaction symmetry. To probe this question even more thoroughly, the populations of VAL and THR are compared in STs and ETs across all four SCOP classes. VAL and THR share (approximately) the same shape and volume, so much so that it constitutes a non-trivial task to distinguish the two, even in a high resolution crystal structure. VAL has a branched hydrocarbon side chain. THR resembles VAL, but with a hydroxyl group in place of one of the VAL's methyl groups, which makes THR comparatively hydrophilic. Four in most-favored 25 STs and five in least-favored 25 STs are found to contain VAL, while not a single most-favored or least-favored STs are found to accommodate THR, even though it has (roughly) the same size as that of VAL and even though it is a bit more hydrophilic than VAL. – All these observations merely reinstate that tunnel properties cannot be mapped to stereochemical and biophysical/biochemical properties of single residues.

### *3.6.4.3. Are more-reactive residues favored in STs and ETs?*

Do structural and evolutionary constraints favor symmetric interactions between residues that are more reactive than others? GLU has one additional methylene group in its side chain than does ASP. The pKa of the $\gamma$-carboxylic group of GLU is higher than the pKa of $\beta$-carboxylic group of ASP. Present set of results show that GLU is found in 11 most-favored STs across all four structural classes, whereas only 6 instances of ASP could be found in most-favored STs. But this observation does not necessarily imply that more-reactive residues are more favored in tunnel populations, a comparative study of CYS and SER population in tunnels exemplifies this point. CYS differs from SER by a single atom; the sulfur of the thiol replaces the oxygen of the alcohol. The proton of the thiol of CYS is much more acidic than the hydroxylic proton of SER, making the nucleophilic thiol(ate) much more reactive than the hydroxyl of SER. However, in an anti-thesis to 'more-reactive more-favored in tunnels' expectation, it is found that SER occurs as components of two most favored STs across four structural classes, but not a single instance of CYS could be found in any of the most-favored STs in any of the structural classes. On the other hand, CYS occurrence is detected in 10 cases as part of lowest-favored STs across structural classes, whereas SER is not found as a part of any of the lowest-favored STs. Equally perplexing can be the observations made from comparative study of PHE, TYR and TRP, in the present context. – These examples simply re-establish the assertion that though residue-interaction-symmetry stems from residue interactions, it is truly a (strong) emergent property; the characteristics and effects of which can only be studied at the level of protein tertiary structures and can neither be mapped nor deduced from properties of individual residues.

*3.6.5. The curious case of CYS in tunnels:* CYS is difficult to be substituted by other residues. Barring TRP, CYS is known to be the most conserved residue in protein structures. – These observations are routinely (and probably correctly) explained by citing CYS's ability to form di-sulphide bridges in proteins. Given the pivotal importance of CYS residues in a protein, it seems paradoxical to note that CYS does not form a component of any of most-favored STs across any of the four SCOP classes and in fact, CYS is found to have maximum number of occurrences among the lowest-favored STs. – This paradox, however, can easily be resolved by recognizing that conserved character of any residue does not necessarily imply symmetry-promoting and symmetry-conserving character of it. Though most-conserved, the global stoichiometric proportion of CYS and TRP in proteins are the lowest (only 1.9% and 1.4%) respectively). This explains why an individual CYS (or TRP) residue can have profound roles in structural organization, but at the same time can account for very little symmetry in pairwise-interaction profiles that it features in.

## 4. Discussion:
*4.1. Comparing tunnels with other works on continuity in protein structural organization:* To get back to the original motivation of the study, that is, to probe whether protein structural organization is discreet or continuous, we compare and contrast our approach with some of the previous approaches that studied the same question. Motivation of many previous works was to detect structural units that lie underneath the organizational level of structural domains; whereby, starting with an attempt to characterize "highly repetitive near-contiguous pieces of polypeptide chain"(Shindyalov and Bourne, 2000), one work identified the (minimally) independent folding units (Tsai et al., 2000) and another reported the consistency in folded nature of ~30-residue loops (Berezowski and Trifunov, 2001). The same philosophy, from a slightly different angle, had motivated (Unger et al., 1989) to search for canonical sub-structures in proteins; later on, (Kolodny et al., 2002), (Tendulkar et al., 2004) and (Szustakowski et al., 2005) identified certain recurrent polypeptide fragments that serve as elementary building blocks. However, insightful as these studies were, it didn't always become clear from their results that whether the origin of the reported (autonomous) sub-structures is due to evolutionary constraints, or is it due to the structural constraints, or due to interplay between these two. Furthermore, a necessary (but not always sufficient) condition to assert continuity in structural organization, viz. detection of a family of continuous bottom-up threads that connect the particular preferences in local interactions to globally defined evolutionary role – could not always be found from the aforementioned works. The STs and ETs (acceptable, unacceptable alike) do not attempt to investigate the (possibly) continuous nature of protein structural organization by detecting newer sub-structures, but establish continuity in the threads that connect the most-elementary unit of structural organization (viz., residue-residue interactions) to its highest level (viz., that of the functional state of the quaternary structures).

***4.2. Why we had to study the common ground among structural and evolutionary constraints:*** As structural unit, a SCOP domain's character is both evolutionary and non-evolutionary. Therefore, a general algorithm to estimate the extent of latent convergence among structural units must necessarily consider the feasible templates satisfying both structural constraints (which ensure appropriate geometrical features so that the statics of mechanical load distribution in a protein is optimized) as well as evolutionary constraints (which ensure that a protein in thermodynamically acceptable state is capable of functioning under the thermodynamical context of a biochemical pathway). Importance of **Table-5** lies precisely therein; this meeting ground between STs and ETs details the degree of universality in acceptability or unacceptability of residue-residue interaction symmetries. Since evolution-driven morphing of one SCOP fold into another has already been reported (Taylor, 2007; Goldstein, 2008), an analysis of progression of tunnel features (the magnitude of CD, tunnel strength, etc.) during fold transformation may bring to light the scope and depth of the influence of cellular-level contexts on the most elementary unit of protein structural organization.

***4.3. Why only a small subset of residues is critical for stability of a functioning protein?*** It is known that only a small number of residues is "critical" for folding of a protein (Vendruscolo et al., 2001; Shimada and Shakhnovich, 2002). It is also known that only with minimal change in residues two sequences with 88% identity can be made to adopt completely different folds (He et al., 2008). Shifting from the paradigm of individual residues to that of pairwise-interactions, it is known that there are statistically significant preferences in them as well (Gregoret and Cohen, 1990; Karplus and Shakhnovich, 1992; Levitt et al., 1997; Keskin and Bahar, 1998; Das and Baker, 2008; Berka et al., 2010). The present work generalizes this observation by reporting that out of 210 possible pairwise-interaction symmetries that connect local-interactions to global-organization, only a chosen few are actually 'accepted' by different structural classes (***Section 3.1-to-Section 3.4***). Notably, a high percentage of interactions that are found to be highly symmetric (and therefore, highly 'acceptable') in one structural class, is found to be highly symmetric (and therefore, highly 'acceptable') in other structural classes too. In addition, the present study reports definite preferences in nature's disliking for certain pairwise-interactions, the symmetries of which are detected to be universally low (and therefore, highly 'unacceptable') in proteins belonging to different structural classes. – Thus, the present work generalizes the previous findings in establishing that, while all residue-residue interactions are important, only a small subset of interactions are actually important to meet geometric/structural and evolutionary constraints in proteins belonging to different structural classes.

**5. Conclusion:**
If protein structural organization is continuous, one will be able to connect the local-scale residual interactions to the global-symmetry of geometric regularities observed in the functional state of quaternary structures. To detect the (possible) continuous lineage that connects patterns in local interactions to that of global structural organization, a different approach was implemented in the present work. It was found that there exists a limited set of latent patterns in residue-residue interaction symmetries that remain invariant across proteins belonging different SCOP classes. Contrary to the expectation to detect preference of any SCOP-class for a definite set of pairwise-interactions, it was found that pairwise interactions that are favored in one structural class are often favored by other structural classes. To what extent, these latent pairwise symmetries in different SCOP classes are dependent upon evolutionary constraints – was investigated too. It was found that principles for emergence of quaternary structural organization do not correlate well with the principles for emergence of secondary structural organization.